\documentclass[a4paper, journal, 11pt]{IEEEtran}
\usepackage{cite}
\usepackage{amsmath,amssymb,amsfonts}
\usepackage{algorithmic}
\usepackage{graphicx}
\usepackage{textcomp}
\usepackage{xcolor}
\usepackage{acronym}
\usepackage{array, makecell}
\usepackage{multirow}
\newacro{AFM}{antiferromagnetically}
\newacro{LIF}{leaky integrate-and-fire}
\newacro{PMA}{perpendicular magnetic anisotropy}
\newacro{SAF}{synthetic antiferromagnet}
\newacro{FM}{ferromagnet}
\newacro{MTJ}{magnetic tunnel junction}
\newacro{STT}{spin-transfer-torque}
\newacro{DW}{domain wall}
\newacro{HM}{heavy metal}
\newacro{LLGS}{Landau-Lifshitz-Gilbert-Slonczewski}
\newacro{SOT}{spin-orbit torque}
\newacro{STT}{spin-transfer torque}
\newacro{TL}{top layer}
\newacro{BL}{bottom layer}
\newacro{DMI}{Dzyaloshinskii-Moriya interaction}
\newacro{CIP}{current-in-plane}
\newacro{CPP}{current-perpendicular-to-plane}
\newacro{SNN}{spiking neural network}
\newacro{LTP}{long term potentiation}
\newacro{LTD}{long term depression}

\begin{document}

\title{Design of Spintronics-based Neuronal and Synaptic Devices for Spiking Neural Network Circuits\\
{\small(Invited Paper)}
}

\author{Debasis Das, Yunuo Cen, Jianze Wang, Xuanyao Fong\\	
	\thanks{The authors are with the Dept. of Elect. \& Comp. Engineering, National University of Singapore,
		Singapore, 117583.}
}
\maketitle

\begin{abstract}
Topologically stable magnetic skyrmion has a much lower depinning current density that may be useful for memory as well as neuromorphic computing. 
However, skyrmion-based devices suffer from the Magnus force originating from the skyrmion Hall effect, which may result in unwanted skyrmion annihilation if the magnitude of the driving current gets too large.
A design of an artificial neuron and a synapse using a synthetic antiferromagnetically coupled bilayer device, which nullifies the Magnus force, is demonstrated in this work. 
The leak term in the artificial leaky integrate-and-fire neuron is achieved by engineering the uniaxial anisotropy profile of the neuronal device. 
The synaptic device has a similar structure as the neuronal device but has a constant uniaxial anisotropy.
The synaptic device also has a linear and symmetric weight update, which is a highly desirable trait of an artificial synapse. 
Neuronal and synaptic devices based on magnetic domain-wall (DW) motion are also studied and compared to skyrmionic devices.
Our simulation results show the energy required to perform such operation in DW or skyrmion-based devices is on the order of a few fJ.

\end{abstract}

\begin{IEEEkeywords}
spiking neuron, synapse, magnetic domain wall, magnetic skyrmion
\end{IEEEkeywords}

\section{Introduction} \label{sec:intro}
\begin{figure}[!t]
    \includegraphics[scale=0.4]{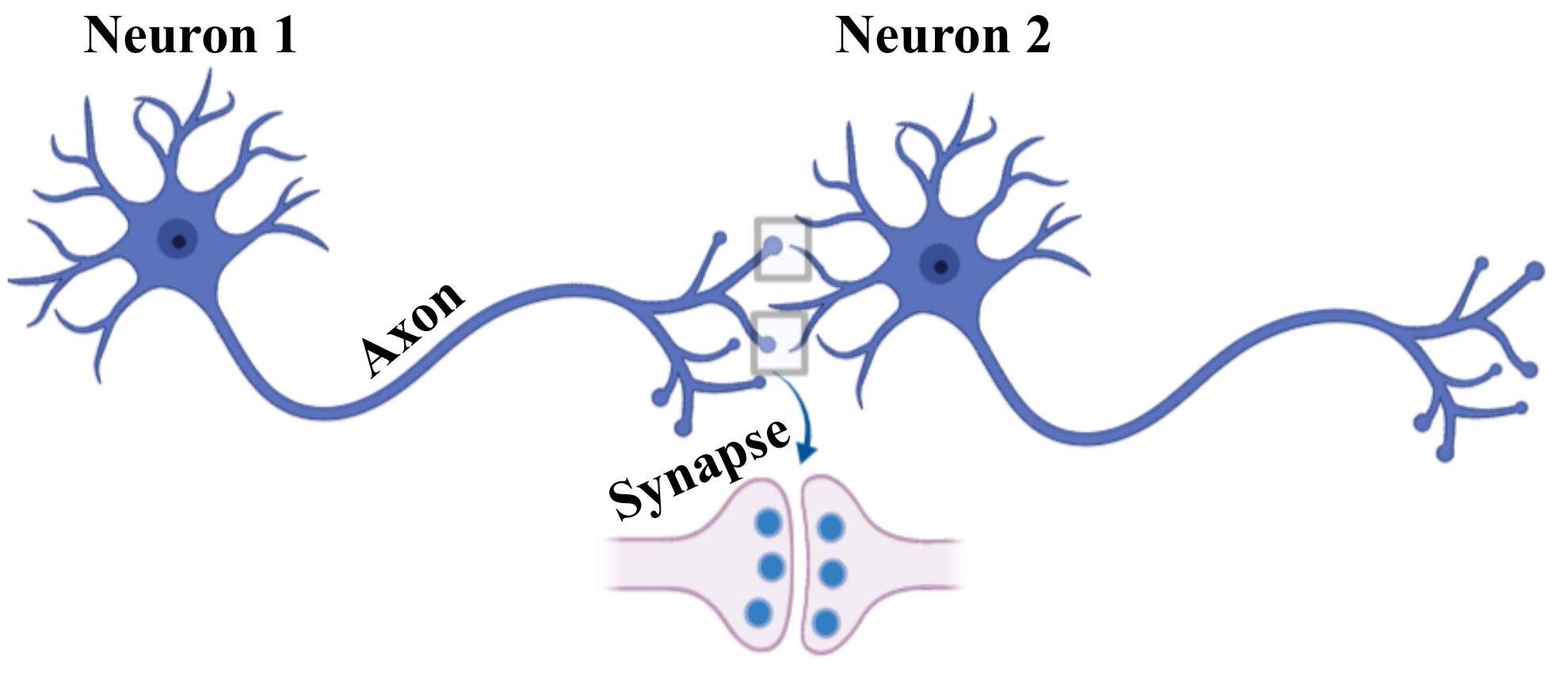}
	\caption{Schematic of biological neurons connected via synapses.}
	\label{Fig:Bio_neuron_synapse}
\end{figure}

Brain-inspired neuromorphic computing has emerged as one of the promising approaches that can perform cognitive tasks~\cite{markovic2020physics,roy2019towards} with human-like accuracy. 
Neurons and synapses are the two key components in a biological \acf{SNN} in which billions of neurons are interconnected via trillions of synapses.
Information in an \ac{SNN} propagates in the form of a spike that is transmitted from one neuron (\emph{pre-neuron}) to the next neuron (\emph{post-neuron}) connected via a synapse---the spikes get modified by the synaptic weight during this process.
The schematic of a simple \ac{SNN} consisting of two neurons that are connected by synapses is shown in Fig.~\ref{Fig:Bio_neuron_synapse}. 
When the pre-neuron (Neuron1) fires a spike, the synapse modifies the spike with its synaptic weight before forwarding it to the post-neuron (Neuron2).

Among various emerging devices (\textit{e.g.}, resistive memory~\cite{yu2011electronic,park2012rram} and phase-change memory~\cite{tuma2016stochastic,kim2015nvm}), spintronics~\cite{sengupta2016proposal,brigner2019shape,zhang2016all,chen2018magnetic} is arguably the most promising technology due to its non-volatile nature, low power consumption, ultrafast dynamics, higher endurance,  and stochastic nature~\cite{zhang2014spintronics,shim2017stochastic,das2021fokker}.
Magnetic \acf{DW}~\cite{parkin2008magnetic, das2022self} and skyrmion-based devices~\cite{sampaio2013nucleation, das2019skyrmion} are the most popular among various other spintronics devices. 
Among these two, skyrmion-based devices are promising due to their topological stability (implies robustness against defects) and lower depinning current density (implies lower power consumption) compared to magnetic domain wall-based devices. 
However, skyrmion-based devices have limitations due to the Magnus force~\cite{nagaosa2013topological,chen2017skyrmion}, which acts perpendicular to the direction of the injected current.
Consequently, the skyrmion moves along a resultant direction determined by spin-torque due to injection current and the Magnus force.
The magnitude of the Magnus force is directly proportional to the velocity of the skyrmion, which is proportional to the magnitude of the injected current density. 
If the magnitude of the injected spin current is too large, the Magnus force may be large enough to annihilate the skyrmion at the nanotrack edge during its motion, which limits the utility of skyrmions for high-speed applications.

To drive the DW/skyrmion in a nanotrack, a spin current is needed, which can be injected either by  (i) \acf{CIP} or (ii) \acf{CPP} scheme. 
CPP scheme stands out to be more energy efficient compared to CIP, due to the lower current density needed to drive the DW/skyrmion. 
Elimination of the Magnus force in a skyrmion-based device was demonstrated using the CIP scheme in Ref.~\cite{li2017magnetic}, through micromagnetic simulation, with the condition that the non-adiabatic spin-transfer-torque factor, $\beta$, needs to be exactly equal to the Gilbert damping constant, $\alpha$.
Hence, this method is sensitive to the material parameter and any mismatch can introduce the undesired Magnus force.

Alternatively, a bilayer device where the magnetization of two \acf{FM} layers are coupled via an \acf{AFM} exchange coupling may be promising for eliminating the Magnus force. 
The Magnus force can be completely nullified if the coupling constant between the magnetic layers is sufficiently large~\cite{zhang2016magnetic}. 
In this work, we use micromagnetic simulations to demonstrate how this bilayer device can be designed to be neuronal and synaptic devices.  
The design of neuronal and synaptic devices using \ac{DW} motion in a traditional monolayer structure is also presented and compared with the skyrmionic device.

\section{Device structure}

\begin{figure}[b!]
	\includegraphics[scale=0.4]{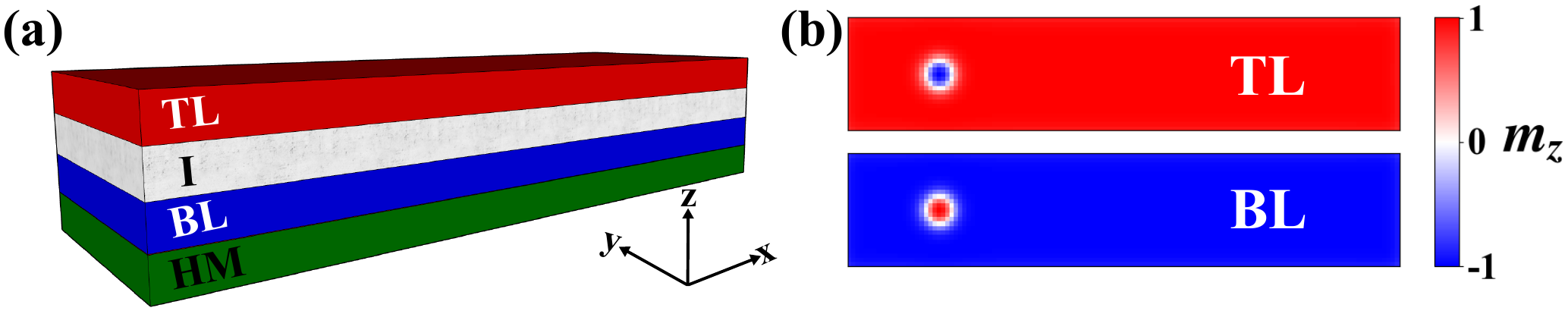}
	\caption{(a) 3D-view of the bilayer device. (b) A two-dimensional color plot of $m_z$ over the $x\mbox{-}y$ plane in TL and BL showing the skyrmions in the device. }
	\label{Fig:Bilayer_device}
\end{figure}

The bilayer system consists of a \ac{HM}/\ac{FM}/insulator (I)/\ac{FM} stack shown in Fig.~\ref{Fig:Bilayer_device}(a).
The \ac{FM} layer adjacent to \ac{HM} is the \ac{BL} whereas the topmost \ac{FM} layer is the \ac{TL}.
Due to the spin-Hall effect, charge current injected through the \ac{HM} layer generates a spin-current that is spin-polarized along the $y$-direction and flows in the $z$-direction into the \ac{BL} to drive the motion of any \ac{DW} or skyrmion in it.
The \ac{HM} layer also gives rise to the \ac{DMI} that stabilizes the skyrmion in the device.
Both \ac{TL} and \ac{BL} are designed with \ac{PMA} and their magnetizations are coupled by \ac{AFM} exchange interaction.
In the following, the initial magnetizations of \ac{TL} and \ac{BL} are assumed to be along +$z$ and -$z$ direction, respectively.

Assume the size of both \ac{TL} and \ac{BL} layers in Fig.~\ref{Fig:Bilayer_device}(a) is 260~nm$\times$50~nm$\times$1~nm in the following discussion.
The membrane potential in the \ac{LIF} neuron device implemented using the bilayer structure in Fig.~\ref{Fig:Bilayer_device}(a) is represented by the position of a skyrmion-pair along the length of the device.
Local injection of a 5$\times10^{14}~\mathrm{A/m^2}$ current pulse that is spin-polarized along the -$z$-direction can nucleate a skyrmion at the desired location of \ac{TL}. 
The magnetization of the core of the nucleated skyrmion points in the -$z$-direction. 
As shown in Fig.~\ref{Fig:Bilayer_device}(b), another skyrmion with opposite magnetization to that in \ac{TL} is simultaneously nucleated in \ac{BL} simultaneously due to the \ac{AFM} coupling.
Input spikes that are integrated by the device are in the form of current pulses that are injected along the length of the device, which moves the skyrmion-pair.

\begin{figure}[!b]
	\includegraphics[scale=0.4]{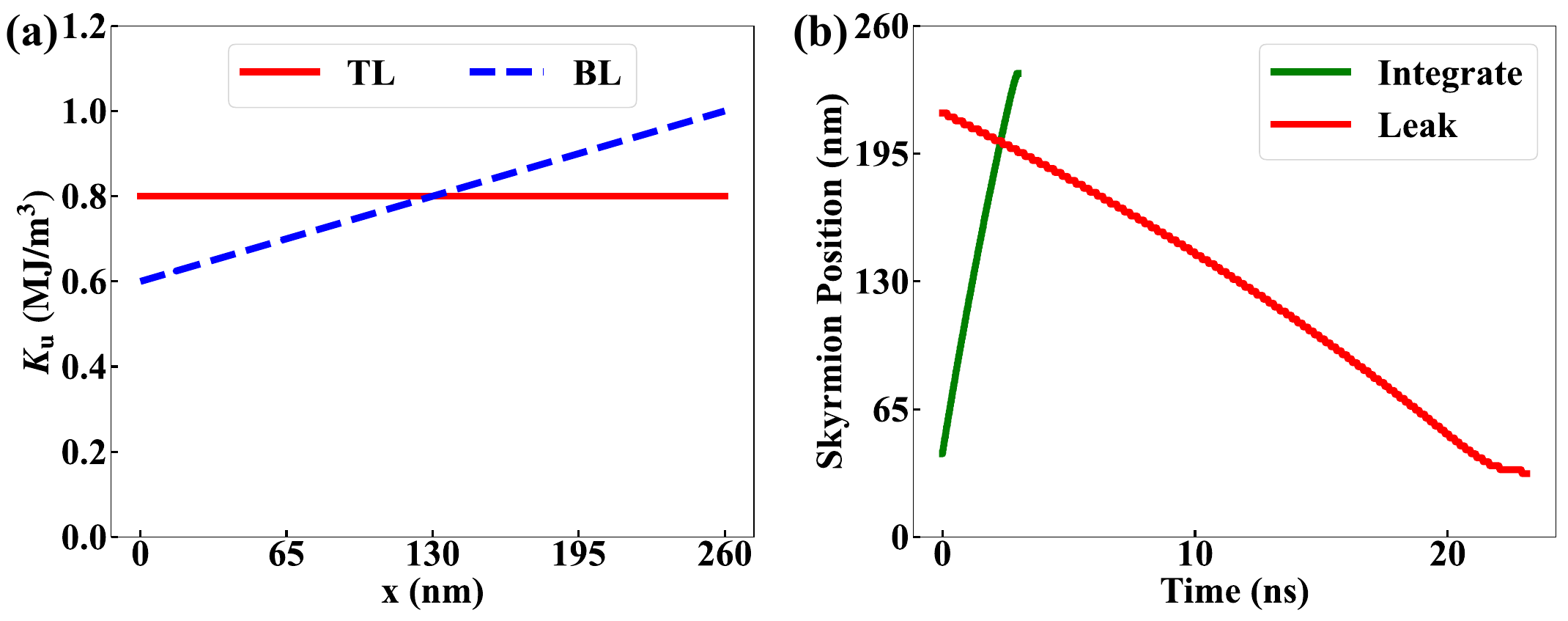}
	\caption{(a)$K_\text{u}$ profile in the bilayer device for implementing leak functionality. (b) Plot for the skyrmion position with time during \emph{integration} and \emph{leak} process.}
	\label{Fig:bi_Ku_plot_leak_int}
\end{figure}

The leak term in the neuron device may be achieved by implementing a gradient in the uniaxial anisotropy constant, $K_\text{u}$, along the length of the \ac{BL}. 
On the other hand, the \ac{TL} has a uniform $K_\text{u}$ as shown in Fig.~\ref{Fig:bi_Ku_plot_leak_int}(a). 
To validate the integration and leak functionalities, we test the device in the following way.
A skyrmion-pair in the \ac{TL} and \ac{BL} is initially nucleated at $x$=50~nm.
A uniform 30~$\times10^{10}\mathrm{A/m^2}$ charge current density is next applied in the \ac{HM} along the +$x$-direction, which moves the skyrmions in \ac{TL} and \ac{BL} in unison.
The position of the skyrmion center in the TL versus time is plotted as the green curve in Fig.~\ref{Fig:bi_Ku_plot_leak_int}(b). 
The skyrmion in the BL has the same behavior due to the \ac{AFM} coupling.
Thus, the position of the skyrmion appears as the time-integral of the injected current and the \emph{integration} functionality of the proposed neuron device is verified.

Next, consider when the skyrmion-pair is initially at $x$=215~nm and no charge current flows through the device.
Notice that, as shown by the red color plot in Fig.~\ref{Fig:bi_Ku_plot_leak_int}(b), the skyrmion in \ac{BL} moves gradually along the -$x$-direction due to the $K_\text{u}$ gradient.
Due to \ac{AFM} coupling, the skyrmion in \ac{TL} moves in unison with that in \ac{BL}.
After 23~ns, the skyrmion stops at $x$=25~nm due to the edge repulsion.
This behavior of the skyrmion resembles the \emph{leak} functionality of neuron.

On the other hand, synaptic devices should behave like a non-volatile memory devices.
This can be achieved in the same bilayer device as the neuron if the $K_\text{u}$ is uniform along the device in both \ac{TL} and \ac{BL}.
The synapse emulates different weight values to modulate the signal transmission from the pre-neuron to the post-neuron. 
Multiple values of the synaptic weight is achieved in a device having multiple conductance levels. 
These conductance levels can be achieved using multiple skyrmions in the device. 
With the application of different current pulses, the skyrmions can be moved between the two sides of the device, which changes the conductance of the detector arising from the magnetoresistance effect.

\begin{figure}[t!]
	\includegraphics[scale=0.43]{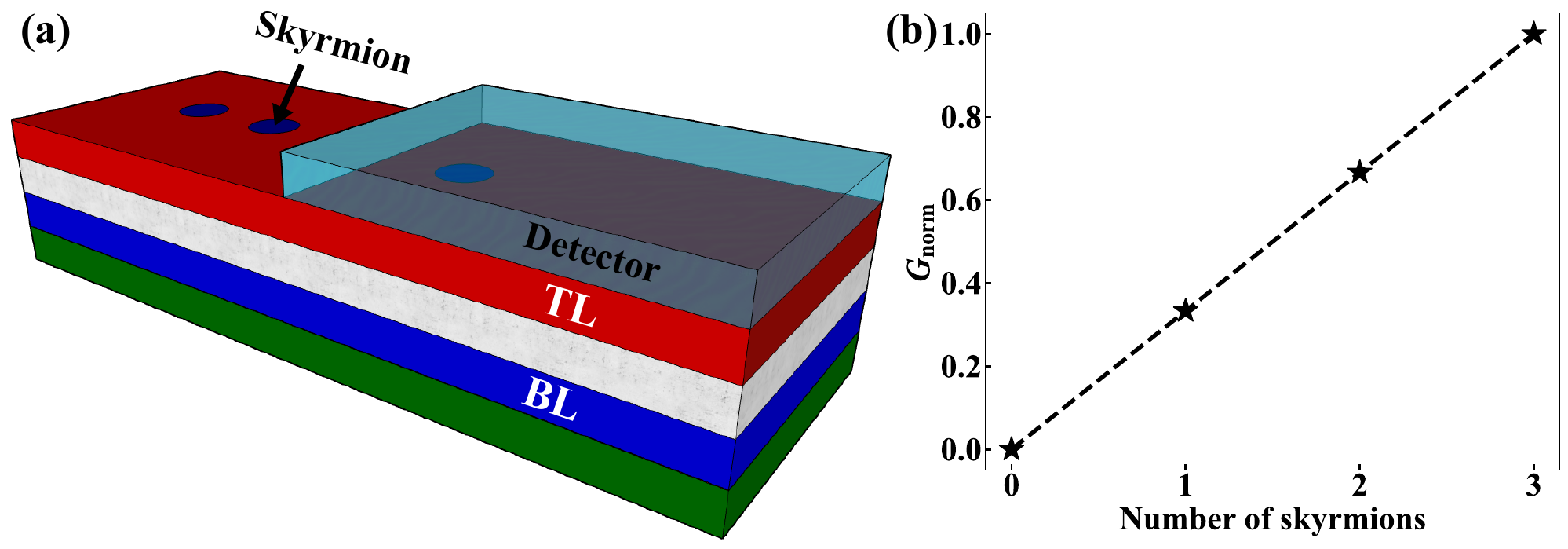}
	\caption{Schematic of a 3D-view skyrmion-based synapse device. (b) Plot for normalized conductance, $\mathrm{G_{norm}}$ of the detector with the number of skyrmions. }
	\label{Fig:Skyrmion_Synapse_demo}
\end{figure}

To demonstrate this, consider if three skyrmions are initially on the left half of a bilayer device and the detector is placed on the other half as can be seen from Fig.~\ref{Fig:Skyrmion_Synapse_demo}(a). 
When current pulses are applied, these skyrmions enter the detector area one by one, which changes the conductance of that area.
The variation of the conductance in normalized scale, denoted by $\mathrm{G_{norm}}$, with the number of skyrmions under the detector area is plotted in Fig.~\ref{Fig:Skyrmion_Synapse_demo}(b) that shows a linear relationship.
Thus, the number of skyrmions under the detector area may be used to represent different weight values in the synapse to modulate the neuronal spikes.

\subsection{Domain-wall (DW) based devices}

\begin{figure}[!b]
	\includegraphics[scale=0.44]{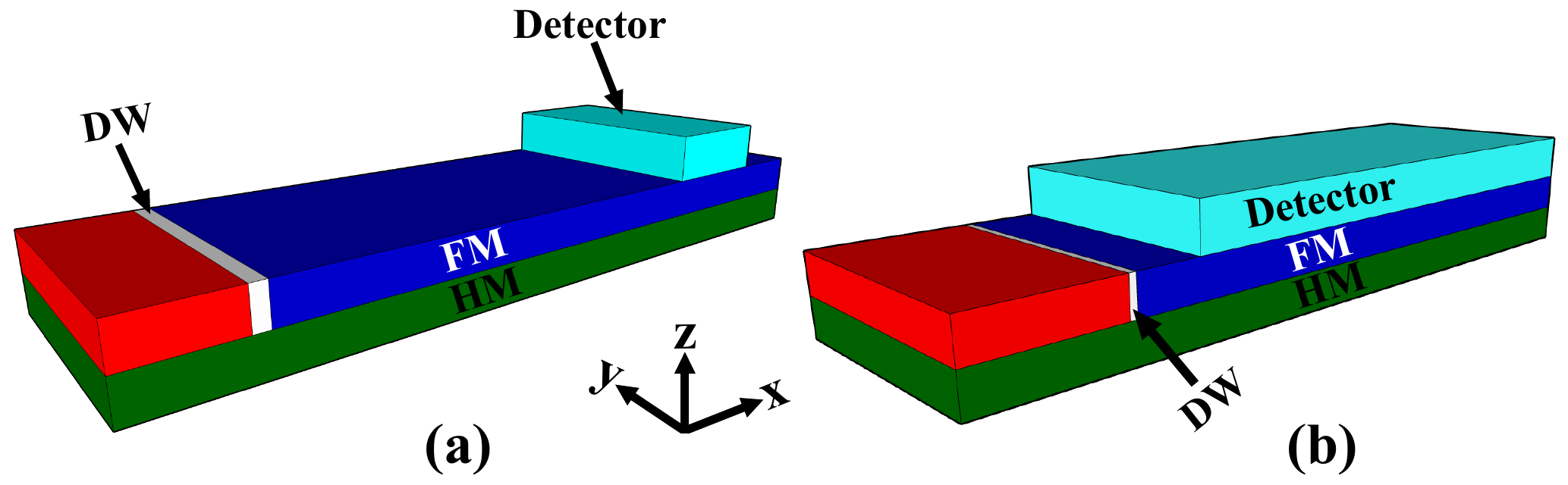}
	\caption{(a) Schematic 3D-view of monolayer \ac{DW}-based (a) neuron and (b) synapse device.}
	\label{Fig:DW_neuron_synapse_demo}
\end{figure}

Magnetic \ac{DW} based devices do not suffer from the Magnus force issue and thus, a conventional monolayer structure may be used to implement these neuronal and synaptic devices.
Consider the structure shown in Fig.~\ref{Fig:DW_neuron_synapse_demo}, which consists of a single \ac{FM} layer with \ac{PMA} and is attached to a \ac{HM} layer.
In the neuron device, the membrane potential is represented by the position of the \ac{DW} shown by the white region between red (magnetized along +$z$) and blue (magnetized along -$z$) regions. 
A $K_{\text{u}}$ gradient is engineered into the \ac{FM} layer to implement the leak behavior of the neuron.
To make the synaptic device, a long detector along the length of the device as shown in Fig.~\ref{Fig:DW_neuron_synapse_demo}(b) is required.
The operation principle of the DW-based devices is similar to their skyrmionic counterparts in that the forward (backward) movement of the \ac{DW} under the detector would increase (decrease) the conductance of the detector.

\section{Results and Discussion}

Magnetization dynamics in \ac{FM} layers is simulated in the Object Oriented MicroMagnetic Framework (OOMMF)~\cite{OOMMF} with the DMI extension module~\cite{DMI_extension}.
In all our simulations, the simulation volume is discretized into 2~nm$\times$2~nm$\times$1~nm cells.
The material parameters assumed are those for $\mathrm{Co\vert Pt}$ system~\cite{sampaio2013nucleation, zhang2016magnetic}.

\subsection{Demonstration of neuron device}

The integration and leak functionality of a bilayer-based skyrmionic neuron device has already been described in Section~\ref{sec:intro}. 
Here, the performance of the skyrmion and domain wall-based neuron devices are explained. 

First, consider the skyrmion-based neuron device. 
The membrane potential of the neuron is represented by the position of the skyrmion core in \ac{TL}.
The firing of the neuron occurs when the skyrmion reaches near the right end of the device, which may be sensed by a change of the local magnetoresistance measured using a detector based on the \ac{MTJ}. 
To achieve the leaky behavior, a $K_\text{u}$ gradient described by $K_\text{u}(x)=K_\text{u}^{c}+x\Delta{}K_\text{u}$ where $\Delta{}K_\text{u}$=1.538 $\mathrm{GJ/m^4}$, $K_\text{u}^{c}$=0.6~$\mathrm{MJ/m^3}$, and $x$ is the position along the length of the nanotrack, is engineered into \ac{BL}.
The $K_\text{u}$ of \ac{TL} is at a constant value along its length.

\begin{figure}[!t]
	\includegraphics[scale=0.44]{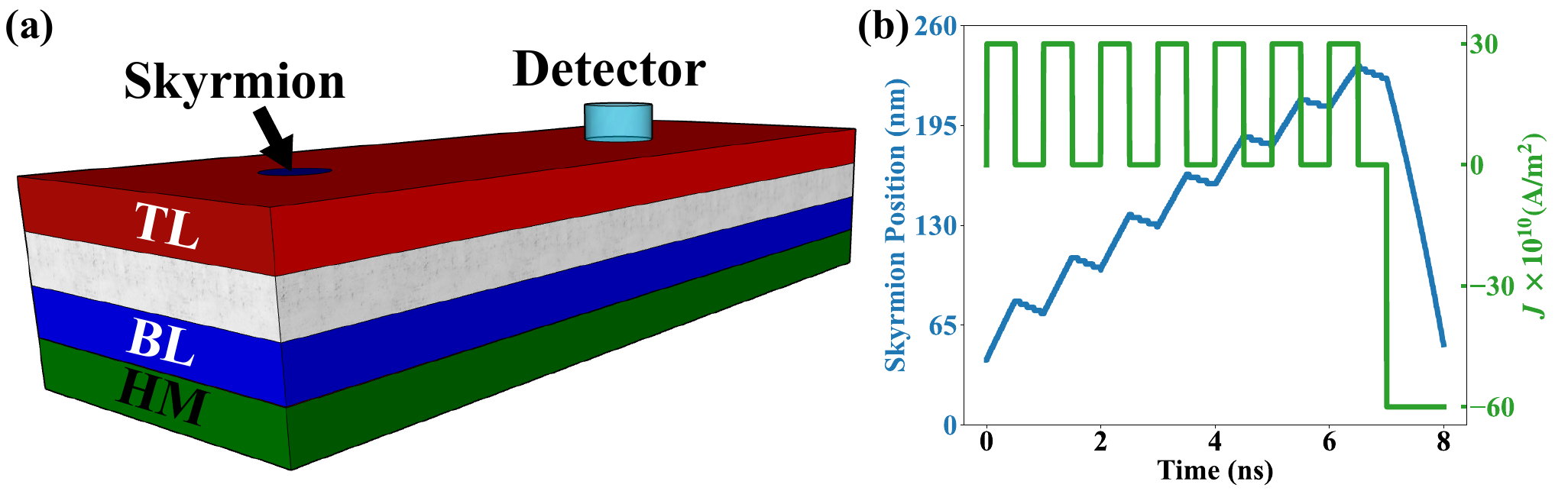}
	\caption{(a) Schematic 3D-view of skyrmion-based neuron device. (b) Variation of the applied input current density (green) and corresponding skyrmion position (blue) along nanotrack length with time. }
	\label{Fig:Skyrmion_neuron_demo}
\end{figure}

Now, consider that the skyrmion was nucleated at $x$=40~nm and the detector is placed at $x$=230~nm, as shown by the schematic in Fig.~\ref{Fig:Skyrmion_neuron_demo}(a). 
A square current density pulse ($J=$30~$\times10^{10}\mathrm{A/m^2}$) having 0.5~ns pulse width and 1~ns period is applied along the +$x$-direction to drive the skyrmion as shown by the green color plot in Fig.~\ref{Fig:Skyrmion_neuron_demo}(b). 
During the high state of the applied pulse, the skyrmion moves in the +$x$-direction.
During the low state, the skyrmion moves in the -$x$-direction due to the $K_\text{u}$ gradient.
The motion of the skyrmion is demonstrated by the blue graph of the skyrmion position versus time in Fig.~\ref{Fig:Skyrmion_neuron_demo}(b).
Once the skyrmion reaches under the detector (at 7~ns), the neuron fires and a reset circuitry (not shown) is triggered to activate a reset 60~$\times10^{10}\mathrm{A/m^2}$ current pulse that resets the skyrmion to its initial position.
This reset pulse is applied for 1~ns, shown in Fig.~\ref{Fig:Skyrmion_neuron_demo}(a), which moves the skyrmion along -$x$-direction as shown in Fig.~\ref{Fig:Skyrmion_neuron_demo}(b) (between 7~ns and 8~ns).
The energy consumption to move the skyrmions in the nanotrack may be calculated as,
\begin{equation}
E=\rho_\text{HM}L_\text{HM}A_\text{HM}J^{2}t_{\mathrm{delay}}
\label{Eq:E_update}
\end{equation}where $\rho_\text{HM}=100~\mathrm{\mu\Omega-cm}$ is the resistivity of the HM~\cite{nguyen2016spin}.
$L_\text{HM}$, $A_\text{HM}$, $J$, and $t_{\mathrm{delay}}$ are the length (along the $x$-axis), cross-sectional area (in the $yz$-plane) of the HM layer, current density, and time delay to move the skyrmion.
The energy required for the rest-to-fire-to-reset operation of this neuron is $~$8.775~fJ.

\begin{figure}[!b]
	\includegraphics[scale=0.45]{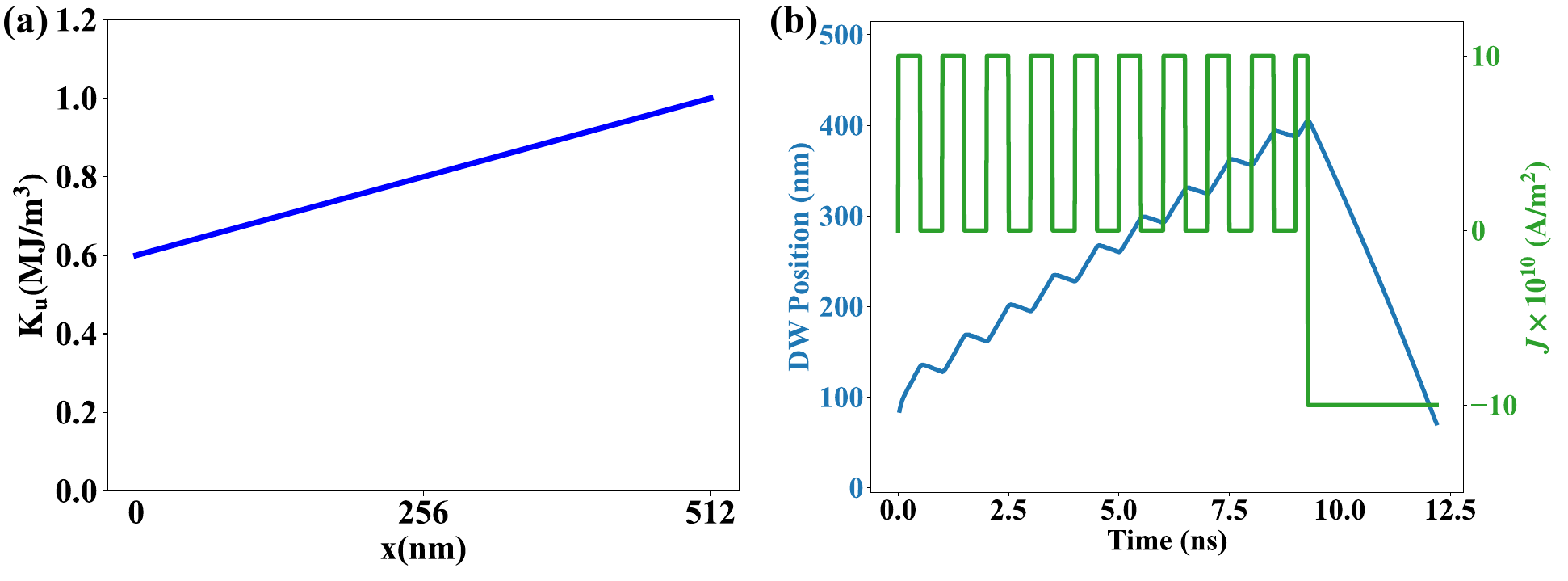}
	\caption{(a) $K_\text{u}$ profile in the \ac{FM} layer for the monolayer device. (b) The plot of the current
		pulse train as the input stimulus is injected into the HM layer, along with the reset current after the neuron fires shown by the green plot, and variation of the \ac{DW} position due to this current is shown in the blue curve.}
	\label{Fig:Mono_DW_neuron}
\end{figure}

Next, consider the neuron operation in a \ac{DW}-based device shown in Fig.~\ref{Fig:DW_neuron_synapse_demo}(a).
Unlike the skyrmion-based device, the lateral dimension of the \ac{DW}-based neuron device is 512~nm$\times$32~nm and the thickness of the \ac{FM} layer is 1~nm. 
A similar $K_\text{u}$ gradient along the length of the device as shown in Fig.~\ref{Fig:Mono_DW_neuron}(a) is engineered into the \ac{FM} to achieve the leak functionality. 
For driving the \ac{DW}, a square charge current pulse ($J=$10$\times10^{10}~\mathrm{A/m^2}$) with a period of 1~ns and pulse-width of 0.5~ns is injected into the \ac{HM} as shown by the green plot in Fig.~\ref{Fig:Mono_DW_neuron}(a). 
Similar to the skyrmion device, the \ac{DW} also moves along the +$x$-direction during a high state of the injected pulse and moves backward during the low state due to the $K_\text{u}$ gradient. 
The variation of \ac{DW} position versus time is shown by the blue plot in Fig.~\ref{Fig:Mono_DW_neuron}(b). 
When the \ac{DW} crosses $x=$420~nm, the conductance state of the detector reaches a maximum, which leads to the firing of the neuron. 
Our simulation results show that it takes 9.26~ns for the \ac{DW} to go from its initial position to the firing position in the detector.
A reset circuitry is needed to generate a current pulse to return the \ac{DW} back to its initial position. 
For reset, the magnitude of $J$ is also 10$\times10^{10}~\mathrm{A/m^2}$ and it takes 2.92~ns to complete the reset process. 
A total of 1.25~fJ energy is consumed to complete the rest-to-fire-reset process (0.779~fJ for the firing process and 0.471~fJ for the reset process). 

\subsection{Demonstration of synapse device}

\begin{figure}[!b]
	\includegraphics[scale=0.43]{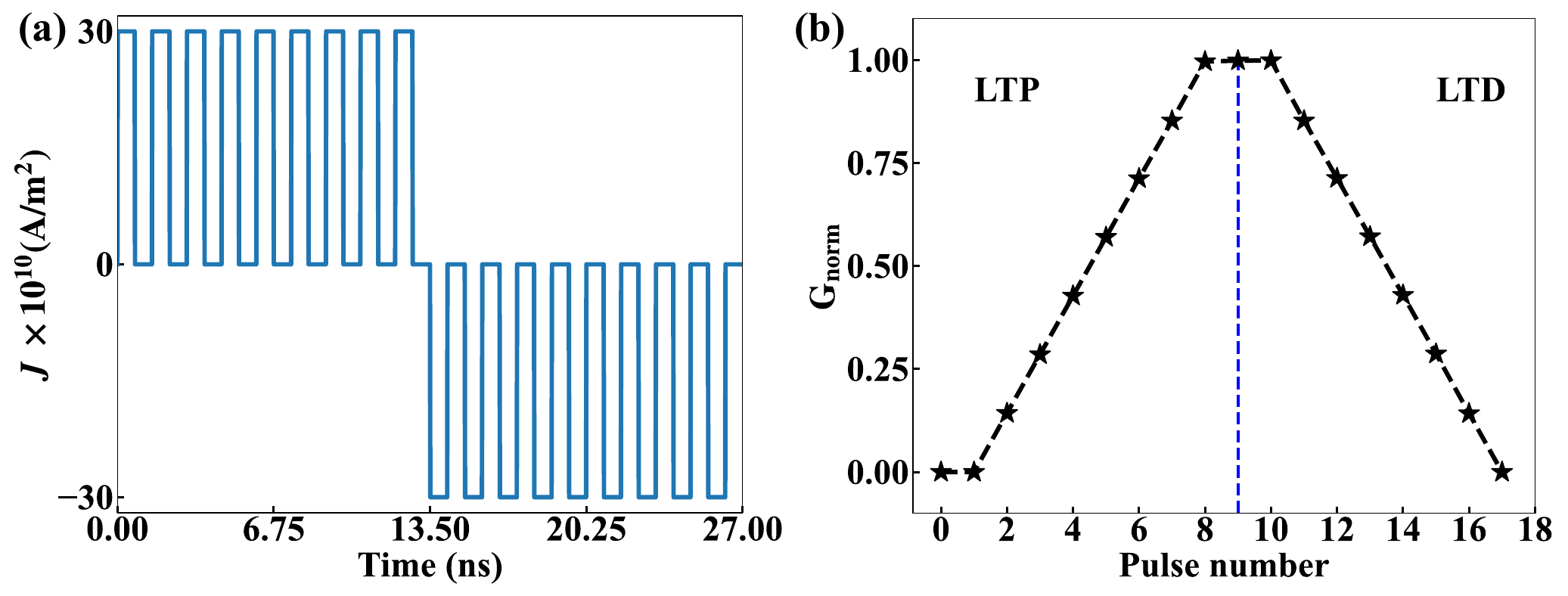}
	\caption{(a) Plot for current density pulse train with time given as an input to a synaptic device. (b) The plot of $\mathrm{G_{norm}}$ vs. pulse numbers during LTP and LTD.}
	\label{Fig:Synapse_device}
\end{figure} 

For the skyrmion-based synapse device, consider a bilayer device shown in Fig.~\ref{Fig:Skyrmion_Synapse_demo}(a) with lateral dimensions of 1000~nm$\times$50~nm, and the thicknesses of both \ac{TL} and \ac{BL} layers are 1~nm. 
As there is no need for the leaky behavior, both \ac{TL} and \ac{BL} have a constant $K_u$ value of 0.8~$\mathrm{MJ/m^3}$. 
The \ac{TL} nanotrack is further divided into two equal halves, where seven skyrmions are nucleated at equidistant on the left half.
An \ac{MTJ}-like detector is on the right half, we place an \ac{MTJ} like a detector.
The operation of the synaptic device is validated by applying a train of the square current pulse into the \ac{HM}.
The magnitude of the current pulse is 30$\times10^{10}~\mathrm{A/m^2}$, with a time-period of 1.5~ns and pulse width 0.75~ns.

The \ac{LTP} operation is achieved by injecting the square current pulses in +$x$-direction whereas the \ac{LTD} is achieved by reversing the direction of current flow. 
The applied current density ($J$) for the \ac{LTP} and \ac{LTD} operation is shown in Fig.~\ref{Fig:Synapse_device}(a).
Individual $J$ pulse applied during the \ac{LTP} operation moves the skyrmions one-by-one from the left half into the detector placed at the right half of the device. 
As the skyrmions move into the detector area, the conductance of the detector starts to increase which is shown in Fig.~\ref{Fig:Synapse_device}(b) by the plot of the normalized conductance, $\mathrm{G_{norm}}$ of the detector versus the number of pulses applied during the \ac{LTP} operation. 
The eight discrete $\mathrm{G_{norm}}$ values marked by black asterisks in Fig.~\ref{Fig:Synapse_device}(b) present the possible synaptic weight for the modulation of the neuron spikes.
These eight weights represent different conductance values since the number of skyrmions in the detector area can vary from 0 to 7.
When all seven skyrmions are in the detector area, the weight/conductance attains a maximum value of 1 (normalized scale).
Further pulse current does not increase the conductance as can be seen from the same $\mathrm{G_{norm}}$ value at pulse number 9 and 10 in Fig.~\ref{Fig:Synapse_device}(b).

During the \ac{LTD} operation, the skyrmions in the detector area go back to the left half one-by-one in a similar fashion, which reduces the conductance of the detector area as can be seen from Fig.~\ref{Fig:Synapse_device}(b). 
Our simulation results show a total of 6~ns is required to complete the entire \ac{LTP} or \ac{LTD} operation.
This generates highly linear and symmetric weight update characteristics (Fig.~\ref{Fig:Synapse_device}(b)), which is a highly desirable condition for an artificial synapse.
Using Eq.~(\ref{Eq:E_update}), the energy required to fully change the synaptic weights of our proposed device from minimum to maximum is 27~fJ.
Since there are a total of seven skyrmions in \ac{TL}, the average energy required to adjust the synaptic weight is 3.587~fJ/unit. 

\begin{figure}[!t]
	\includegraphics[scale=0.45]{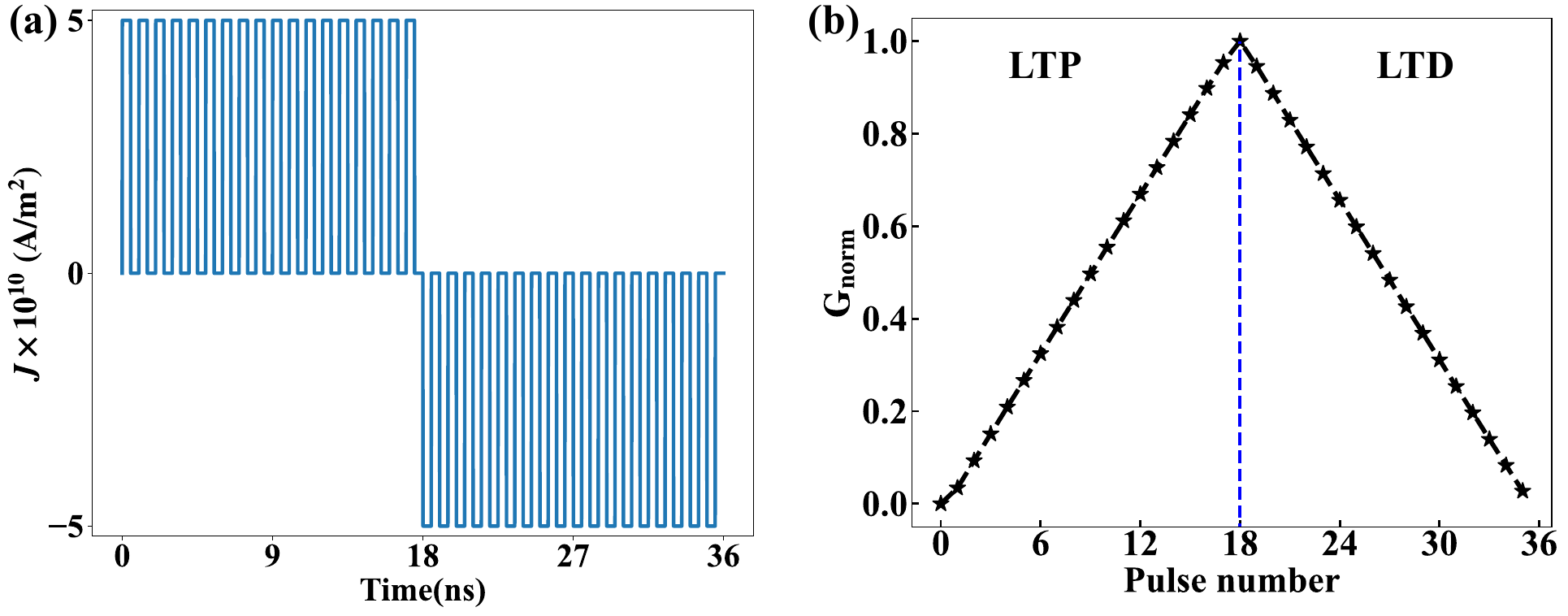}
	\caption{(a) Current density pulse train applied to the \ac{HM} layer for the \ac{LTP} and \ac{LTD} operation in the \ac{DW}-based synaptic device. (b) The plot of $\mathrm{G_{norm}}$ vs. pulse numbers during LTP and LTD.}
	\label{Fig:Mono_DW_synapse}
\end{figure}

Now consider the \ac{DW}-based synaptic device shown in Fig.~\ref{Fig:DW_neuron_synapse_demo}(b). 
Like the previous \ac{DW}-based neuron device, the lateral size of the synaptic device is 512~nm$\times$32~nm, and the thickness of the \ac{FM} layer is 1~nm. 
Unlike, the skyrmion-based synapse device, the detector spans from $x$=40~nm to the rest of the device along its length.
A square current density pulse train with time-period, pulse width, and magnitude of 1~ns, 0.5~ns, and 5$\times10^{10}~\mathrm{A/m^2}$, respectively, is injected into \ac{HM}.
The current is injected for the first 18~ns along the +$x$-direction to perform the \ac{LTP} operation and then in the reverse direction for another 18~ns for the \ac{LTD} operation.
The variation of the injected current density with time for the entire operation is plotted in Fig.~\ref{Fig:Mono_DW_synapse}(a).
The initial position of the \ac{DW} is just outside the detector area.
With the injection of the current pulses during \ac{LTP}, \ac{DW} starts to move in the +$x$-direction.  
As the \ac{DW} enters the detector area and moves farther in, the conductance of the detector increases as seen in the graph of the normalized conductance, $\mathrm{G_{norm}}$, versus the current pulse number in Fig.~\ref{Fig:Mono_DW_synapse}(b). 
During \ac{LTD}, the \ac{DW} moves in the opposite direction and thus, the conductance of the detector starts to decrease as can be seen from Fig.~\ref{Fig:Mono_DW_synapse}(b). 
The results in Fig.~\ref{Fig:Mono_DW_synapse}(b) show that the weight update during \ac{LTP} and \ac{LTD}, is linear in nature and also highly symmetric.
A total of 19 unique conductance states is achieved by different \ac{DW} positions driven by the injected current.
The energy consumed to fully change the synaptic weights of this device from minimum to maximum is 0.36~fJ, which translates to an average of 0.02~fJ/unit.

\begin{table}[t]
	\centering
	\caption{\label{tab:table1}%
		Results summary
	}
	\label{Table:result}
	\begin{tabular}{|c|c|c|c|}
		\hline 
		Technology & Device & Lateral Dimension & Net Energy  \\ 
		\hline 
		\multirow{2}{4em}{Skyrmion} & Neuron & 260$\times$50~$\mathrm{nm^2}$ & 8.775~fJ\\\cline{2-4}
		& Synapse & 1000$\times$50~$\mathrm{nm^2}$ & 27~fJ\\
		\hline
		\multirow{2}{4em}{\ac{DW}} & Neuron & 512$\times$32~$\mathrm{nm^2}$ & 1.25~fJ\\\cline{2-4}
		& Synapse & 512$\times$32~$\mathrm{nm^2}$ & 0.36~fJ\\
		\hline
		
	\end{tabular}   
\end{table}

A comparison of the neuron and synapse devices studied in this work is shown in Table~\ref{Table:result}. 
These results show that \ac{DW}-based device has advantages in lower energy consumption and the same device geometry may be used as a neuron and synapse with just a modification in the $K_\text{u}$ profile.
The lower energy consumption is because the current needed to drive the \ac{DW} motion is 10\% of that needed in the skyrmionic devices.
Both the skyrmionic and \ac{DW}-based synapses produce highly symmetrical and linear weight update characteristics, which is one of the key requirements for an artificial synapse.
However, the number of conductance states is higher in the \ac{DW}-based synaptic device as compared to its skyrmionic counterpart, in which the number of conducting states depends on the number of available skyrmions.
It can be anticipated that the \ac{DW}-based synapse emulates \acp{SNN} with 4-bit weights, which can lead to improved inference accuracy as compared to 3-bit weights in the skyrmionic synapse.
Also, due to the active repulsion force between the skyrmions, a longer device is required to accommodate the large number of skyrmions needed for the number of conducting states.

\section{Conclusions}

In summary, spintronics neuron and synapse device concepts using skyrmion and \ac{DW} were investigated. 
For the neuron devices, the membrane potential is represented by the position of the skyrmion or the \ac{DW}, whereas the synaptic weight is determined by the number of conductance states in a detector area of the synapse device. 
Between \ac{DW} and skyrmion devices, \ac{DW}-based devices seem more advantageous as it consumes 86\% and 97\% less energy for the neuron and synapse devices respectively.
These impressive advantages in energy consumption stem from the 9$\times$ larger current needed for skyrmion motion as compared to \ac{DW} motion.

\bibliographystyle{IEEEtran}
\bibliography{Reference}

\end{document}